\documentclass[doublecol]{epl2}
\usepackage{amsmath}
\usepackage{amssymb}
\usepackage{amsfonts}
\usepackage{graphicx}
\usepackage{graphics}
\usepackage{epsf}
\usepackage{multibib}
\usepackage[T1]{fontenc}
\usepackage{lmodern}
\usepackage[english]{babel}
\usepackage[babel]{microtype}
\usepackage{epstopdf}

\title{ Effective viscosity of non-gravitactic \mbox{\textit{Chlamydomonas Reinhardtii} microswimmer suspensions}}

\author{Matthias Mussler\inst{1}, Salima Rafaï\inst{2}, Philippe Peyla\inst{2} and Christian
Wagner\inst{1}}

\institute{
  \inst{1} Experimentalphysik, Saarland University, Germany

  \inst{2} Laboratoire Interdisciplinaire de Physique - UJF-CNRS, UMR 5588,
Grenoble, France
}


\abstract{
Active microswimmers are known to affect the macroscopic viscosity
of suspensions in a more complex manner than passive particles. For
puller-like microswimmers an increase in the viscosity has been observed.
It has been suggested that the persistence of the orientation of the
microswimmers hinders the rotation that is normally caused by the
vorticity. It was previously shown that some sorts of algaes
are bottom-heavy swimmers, i.e. their centre of mass
is not located in the centre of the body. In this way, the algae affects the
vorticity of the flow when it is perpendicular oriented to the axis of gravity. This orientation of gravity to vorticity
is given in a rheometer that is equipped with a cone-plate geometry.
Here we present measurements of the viscosity both in a cone-plate
and a Taylor-Couette cell. The two set-ups yielded the same increase
in viscosity although the axis of gravitation in the Taylor-Couette cell
is parallel to the direction of vorticity. In a complementary experiment we tested the orientation of the direction of swimming through microscopic observation of single \textit{Chlamydomonas reinhardtii} and could not identify a preferred orientation, i. e. our specific strain of \textit{Chlamydomonas reinhardtii} are not bottom-heavy swimmers. We thus conclude that
bottom heaviness is not a prerequisite for  the increase of viscosity and that the effect of gravity on the rheology of our strain of \textit{Chlamydomonas reinhardtii} is negligible. This finding reopens the question of whether origin of persistence in the orientation of cells is actually responsible for the increased viscosity of the suspension.
}

\begin{document}

\maketitle

\section{Introduction}

Colloidal suspensions are a classical example of complex fluids and
many of the rheological properties are now theoretically described quite accurately.
However, the contribution of active particles to the effective viscosity of a suspension is more complex. Recently,  interest has arisen in the theoretical description of microswimmer suspensions \cite{Hatwalne.2004,Sokolov.2009,Haines.2009,Saintillan.2010,Giomi.2010,Ramaswamy.2012}
but experimental data on their rheology remain limited \cite{Sokolov.2009,Drescher.2010,Rafai.2010}.
Motile micro-organisms, either unicellular or multicellular, can be found
throughout our eco system, and there is interest in the use of active
fluids for many applications, including development of bio fuels, such as cyanobacteria
that produce ethanol \cite{Deng.1999} and microalgae that produce hydrogen \cite{Ghirardi.2000}.

Two types of microswimmers can be distinguished: pushers \textendash{}
which push the fluid behind their body - such as \textit{Escherichia coli} and pullers \textendash{} which pull the fluid in front of their body
 - such as \textit{Chlamydomonas reinhardtii} (CR). Here the notion of "behind" and "in front of" refer to the motion direction of the cells.
It has been recently predicted that slender pushers decrease the effective
viscosity of a suspension that is submitted to shear, whereas slender pullers have
the opposite effect \cite{Hatwalne.2004,Haines.2009,Gyrya.2010,Saintillan.2010}. Rheological characterisations  of CR in a cone-plate geometry
have been previously performed by two of the
authors of the present study \cite{Rafai.2010}. Despite their nearly spherical shape,
the viscosity of a suspension of living CR is significantly increased
compared with a suspension of fixed CR, which can be accurately modelled as a
suspension of passive microspheres of the same diameter. The increase in the viscosity of living CR was attributed to the anisotropy
in the orientation of the cells \cite{Rafai.2010}. Due to their nearly spherical shape, this anisotropy has been suspected to
result either from an external torque due to gravity or from an effective
large hydrodynamic aspect ratio of the cell. A similar comparison was presented in Ref. \cite{Sokolov.2009} in which the bacteria were immobilized by restricting oxygen.

In ref. \cite{Hatwalne.2004} a set of coarse-grained equations that govern the hydrodynamic velocity, concentration and
stress fields in a suspension of active and energy-dissipating solution were
analyzed and several predictions for the rheology of such systems were given.
This model was developed further in another study \cite{Gyrya.2010} in which
a 2D model for a suspension of microswimmers in a fluid was analytically
and numerically analysed in the dilute regime. In Ref. \cite{Gyrya.2010} the analysis has been performed both in dilute
regime (no swimmer-swimmer interactions) and numerically in a moderate
concentration regime (with swimmer-swimmer interactions). The effect of decreasing
viscosity for pusher particles has been described, e.g., in ref. \cite{Giomi.2010}
and the theoretical results were compared with experiments
with suspensions of \textit{Bacillus subtilis} \cite{Haines.2009}. In ref. \cite{Saintillan.2010} an orientation distribution within a
shear flow was used to characterise the configuration of a suspension
of microswimmers. The bodies of the microswimmers were modelled by a force dipole
model that is carried by a slender body.

In ref. \cite{Durham.2009} it was shown that the algae \textit{Chlamydomonas nivalis} are bottom-heavy and orient in the gravitational field. In this article, we provide complementary experiments that aim to determine the effect of gravity on the effective viscosity of a \textit{Chlamydomonas reinhardtii} suspension. Our specific strain of CR had a nearly spherical shape and was not explicitly known as a bottom-heavy swimmer. We measured the effective viscosity of a CR suspension in a Taylor-Couette
geometry (fig. \ref{fig:2}) at different concentrations. In this geometry, gravity is
oriented in the same direction as the vorticity of the flow. However, we
found the same changes in viscosity that were observed in a cone-plate geometry (fig. \ref{fig:1}). This result
shows that gravity does not play a role in the effective viscosity
of our \textit{Chlamydomonas reinhardtii} suspensions.

\section{Theoretical description}

\begin{figure}
\onefigure[width=8.5cm]{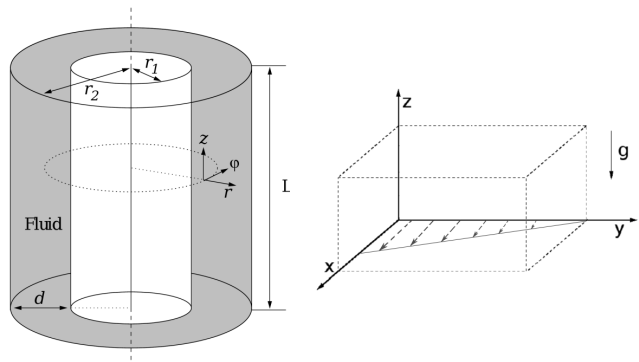}
\caption{
Left: Schematic sketch of the Taylor-Couette System. The outer cylinder
is fixed and the stress is exerted by the inner cylinder, which is
driven by the rheometer. Right: The orientation of the shear gradient in the Taylor-Couette geometry.
The arrows show the shear in the x-y layer for a Taylor-Couette geometry in which the axis of vorticity is parallel to gravity }
\label{fig:2}
\end{figure}

\begin{figure}
\onefigure[width=8.5cm]{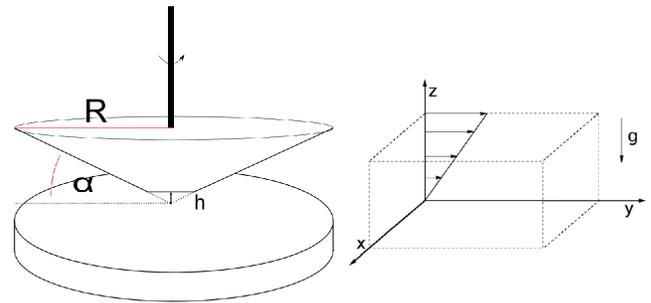}
\caption{Left: Schematic sketch of the cone-plate system. The plate
is fixed and the stress is exerted by the cone, which is
driven by the rheometer. Right: The orientation of the shear gradient in the cone plate geometry.
The arrows show the direction of the shear in the y-z layer. The axis of vorticity (here x-axes) is perpendicular to gravity.}
\label{fig:1}
\end{figure}

The Einstein relation gives the effective viscosity $\eta_{eff}$ of a
dilute suspension of rigid spherical particles:
\begin{equation}
\label{eq:1}
\eta_{eff}=\eta_{0}(1+\alpha\phi)
\end{equation}

for a solvent viscosity $\eta_{0}$, a volume fraction $\phi$
and an intrinsic viscosity $\alpha$. The latter is defined as the
contribution of the solute to the viscosity of the suspension:

\begin{equation}
\label{eq:2}
\alpha=\lim_{\phi\rightarrow0}\frac{\eta_{eff}-\eta_{0}}{\eta_{0}\phi}.
\end{equation}

For passive rigid spherical particles, the intrinsic viscosity is
$\alpha = 2.5$ \cite{W.Pabst.}. Suspensions that have a given particle size distribution
and a fixed volume fraction of the particle phase behave quantitatively
similarly to the monodisperse suspension described above. Eq. \ref{eq:1} is exact in the limit of dilute suspensions (i.e. for
$\phi\rightarrow0$) \cite{Dorr.2012b}. For more concentrated suspensions,
Krieger and Dougherty's law

\begin{equation}
\label{eq:3}
\eta_{eff}=\eta_{0}\left(1-\frac{\phi}{\phi_{m}}\right)^{-\alpha\phi_{m}}
\end{equation}
has been shown to describe the rheology of rigid sphere
suspensions fairly accurately \cite{KRIEGERI.M..1959} ($\phi_{m}$ is the maximal packing fraction).

Our measurements were performed using a rotational rheometer for both the
cone-plate and the Taylor-Couette geometries. The effective viscosity
was obtained by the following equation:

\begin{equation}
\label{eq:4}
\eta_{eff}=\frac{\tau}{\dot{\gamma}}
\end{equation}
where $\tau$ is the stress and  $\dot{\gamma}$ the shear rate. The shear
rate can be chosen and the stress is measured using the controlled shear
rate mode of the rheometer. In the cone-plate geometry,
gravity is perpendicular to the direction of the flow and the vorticity, whereas
gravity is parallel to the vorticity in a Taylor-Couette geometry,
as shown in fig. \ref{fig:2}and \ref{fig:1}.

\section{Materials and Methods} 

CR can be found worldwide in soil or brick water and can grow in a
medium that lacks carbon and energy sources, if it is illuminated through
a day-night cycle \cite{Gorman.1965}. This organism has a nearly spherical shape
with a diameter of $5-10 \mu m$, a large cup-shaped chloroplast,
a large pyrenoid, and an eyespot that
senses light and causes phototactic behaviour. CR has two flagella
and propels itself using a breast-stroke like movement at a frequency
of approximately $30 Hz$ for the strain used in this study, which results in a
velocity of approximately $50\mu m/s$ as measured in Ref \cite{Garcia.2011}.
Because of its small size and weight and the low propulsion velocity force
compared to the viscous forces, CR is a self-motile low-Reynolds-number
swimmer ($\mathrm{Re=2.5\times10^{-4}}$); hence, inertia is negligible
compared to viscous forces. The Reynoldsnumber is
\begin{equation}
\label{eq:5}
Re=\frac{\rho vd}{\eta_{0}}
\end{equation}
where d is the cell diameter, v the velocity and ${\eta}$ the dynamic viscosity of the surrounding
fluid $(\eta_{0}=1.15 mPas)$.
Wild-type strains were obtained from the
IBPC Lab (Physiologie Membranaire et Molculaire du Chloroplaste, UMR
7141, CNRS et Université Pierre et Marie Curie (Paris VI)). Synchronous
cultures of CR were grown in a Tris-acetate phosphate medium (TAP)
using a 12/12 hr light/dark cycle at $25^{\circ}C$ \cite{Gorman.1965}.
The fluid cultures were typically grown for three days under fluorescent
light before the cells were harvested for the experiments. These cultures
were concentrated to a typical volume fraction of 20-40\% by centrifugation
for 5 minutes at 200g and resuspended in fresh culture medium
to achieve the required volume fractions. The volume fractions
were measured using a Neubauer counting chamber. Samples of different
volume fractions were prepared and the effective viscosity was measured
for both live and fixed cells. The fixed cells were treated with formaldehyde
(2\%) to disable the flagella but to preserve the protein structure.
The cell motility was checked before and after the rheological measurements.
We used a Haake Mars 2 rheometer (Thermo Scientific) equipped with
a cone-plate geometry (cone angle $\alpha=2{^{\circ}}$, radius $R=30mm$,
maximum distance between the cone tip and the plate $\mathrm{h}=105\mu m$)
or a Taylor-Couette geometry (inner cylinder radius $r_{1}=10mm$,
$d=r_{2}-r_{1}=1mm$, height $L=72mm$)) (fig. \ref{fig:2} and and \ref{fig:1}).
The measurements were obtained at $T = 25^{\circ}C$. The viscosity of the diluted suspensions of
CR was on the order of the viscosity of water, which means that that the measurements were obtained at the resolution limits of the rheometer.  Therefore, the viscosity data
were integrated over one complete revolution to prevent the influence of any
imbalances in the air bearing of the rheometer.

For the microscopic measurements, the cell suspension was placed in a sealed capillary of size $0.2 \times 2 \times 5$ mm. The observations were performed using a bright-field microscope with a magnification of $\times 1.25$. The microscope was tilted down on the side to ensure that the direction of gravity was in the observation plane. A red filter was used to avoid any bias due to phototaxis. IDL (Interactive Data Language)-based tracking routines were used to locate the cells in time and to measure their velocities.

\section{Results}

\subsection{Cone-plate geometry}

\begin{figure}
\onefigure[width=8.5cm]{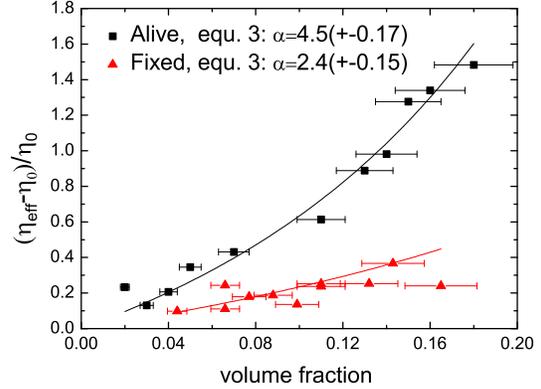}
\caption{The normalised effective viscosity of suspensions of different volume fractions of CR measured in a cone-plate geometry. The squares and triangles indicate the living and fixed CR, respectively. The straight lines are fits using eq. \ref{eq:3} to obtain the intrinsic viscosity, $\alpha$. }
\label{fig:3}
\end{figure}

We first repeated the experimental results that were previously obtained by Rafaï et al.\cite{Rafai.2010}. \revision{In agreement with this preceding investigation a shear thinning of $\eta_{eff}$ can be observed. As discussed in ref. \cite{Rafai.2010}, it reveals competition between the shear rate and cells orientation in the flow. In order to investigate the dependence of effective viscosity on the volume fraction of the suspension, we choose a shear rate of  $\dot{\gamma}=5\frac{1}{s}$}. Fig. \ref{fig:3} shows the relative effective viscosity $\left(\eta_{eff}-\eta_{0}\right)/\eta_{0}$
of live and fixed cells as a function of the volume fraction $\phi$ measured in a cone-plate geometry. In both cases, the relative viscosity increases with increasing volume fraction, but the effect is more pronounced with living cells. More quantitatively, the data were fitted with eq. \ref{eq:3}. We maintained $\alpha$ as a free parameter and chose $\phi_{m}=0.63$. We found values of $\alpha=4.5 \pm0.17$ for living and $\alpha=2.4\pm0.15$ for fixed CR. Moreover, the shearthinning behaviour is also retreived. These results are in agreement with previous work \cite{Rafai.2010}.

\subsection{Taylor-Couette geometry}

\begin{figure}
\onefigure[width=8.5cm]{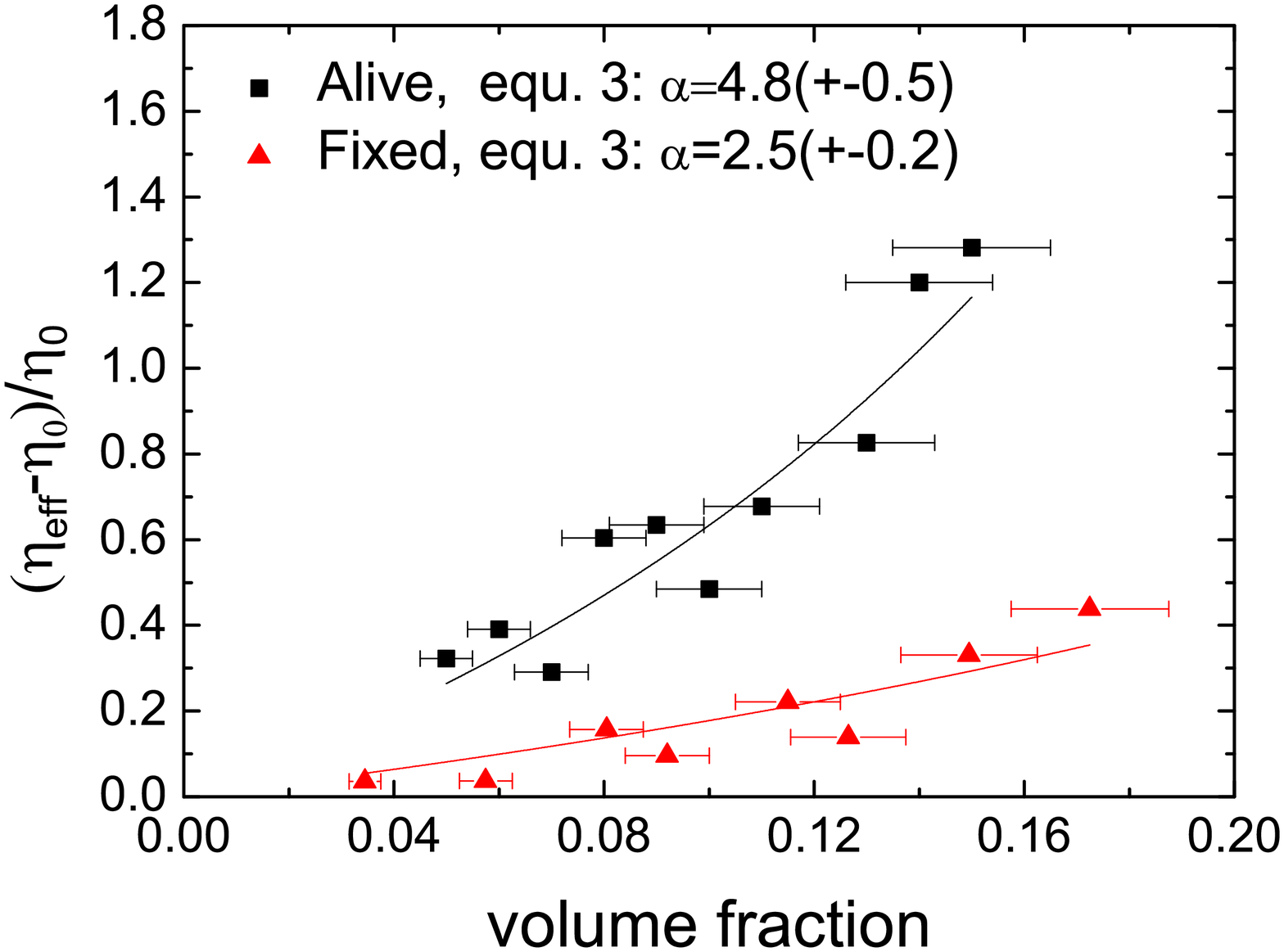}
\caption{The normalized effective viscosity of suspensions of different volume
fractions of CR measured in the Taylor-Couette geometry. The symbols are the same as those used in fig. \ref{fig:3}.}
\label{fig:4}
\end{figure}

To study the influence of gravity on the swimming behaviour of CR, measurements were obtained using a Taylor-Couette geometry. Fig. \ref{fig:4} shows the relative effective viscosity $\left(\eta_{eff}-\eta_{0}\right)/\eta_{0}$ of live and fixed cells as a function of the volume fraction $\phi$ measured at a shear rate of $\dot{\gamma}=5\frac{1}{s}$. The data obtained with both the fixed and living CR were fitted with eq. \ref{eq:3}. As in the previous analysis, we maintained $\alpha$ as a free parameter and chose $\phi_{m}=0.63$. We found $\alpha=4.8$ and $\alpha=2.5$ for live and fixed CR suspensions, respectively. Thus, for both geometries, the intrinsic viscosity of the living and dead CR were identical, within the experimental error.

\subsection{Microscopic measurements}

Some type of algae microswimmers are known to be bottom-heavy, i.e. their mass is not homogeneously distributed. This results in a biased swimming of the cell because the flagella are likely to point upward. However, we tested whether this was the case for the strain used in this study (as well as in Ref. \cite{Rafai.2010}).

\begin{figure}
\onefigure[width=8.5cm]{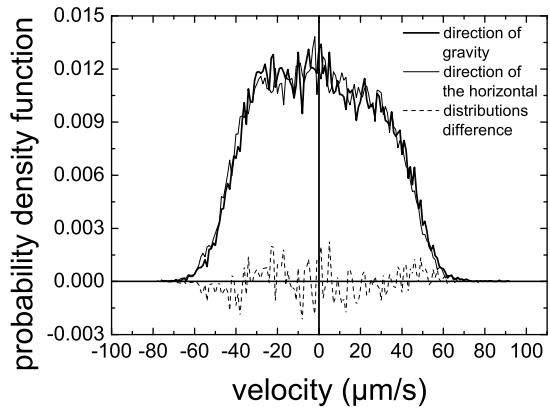}
\caption{The microscopic measured distribution of the velocities of the swimmers along the axis of gravity and along the horizontal axis.}
\label{fig:7}
\end{figure}

\begin{figure}
\onefigure[width=8.5cm]{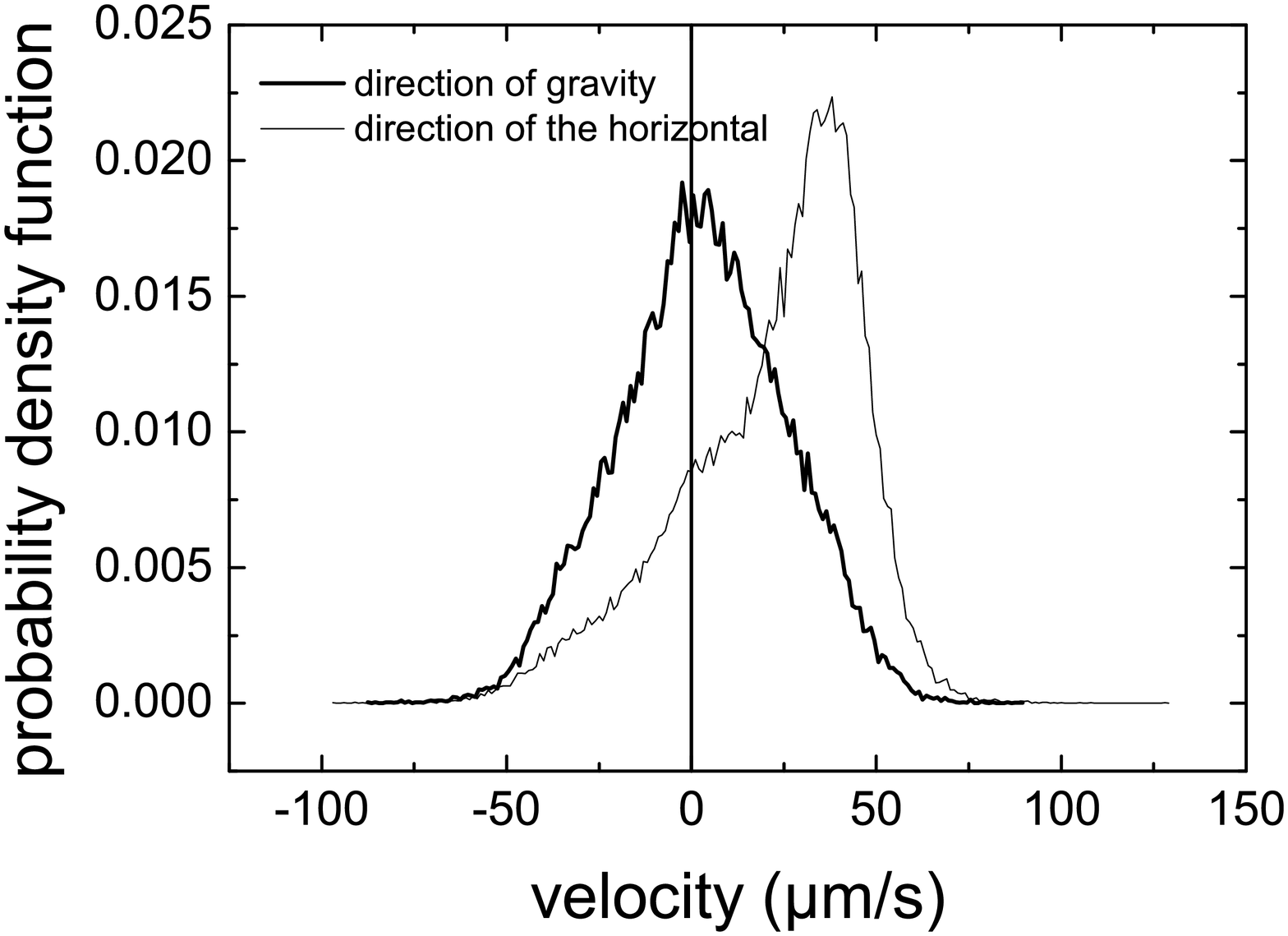}
\caption{The microscopic measured distribution of the velocities of the swimmers along the axis of gravity and along the horizontal axis when the sample was illuminated from the right side.}
\label{fig:8}
\end{figure}

We measured the distribution of velocities of hundreds of cells over twenty minutes both in the direction of gravity and parallel to the horizontal (designated x-direction) (see fig. \ref{fig:7}). We observed no significant difference between the two distributions \revision{and we found only a slight asymmetry, but it was negligible when compared to real taxis.} Thus, the swimming of the cells does not exhibit any bias toward the upward direction, which suggests that this strain is not bottom-heavy. For comparison, when a light was placed on the right side (x>0) of the capillary (fig. \ref{fig:8}), we observed a pronounced biased swimming behaviour, which resulted in an asymmetric distribution of the velocities toward the x>0 region.

\section{Discussion}
\begin{figure}
\onefigure[width=8.5cm]{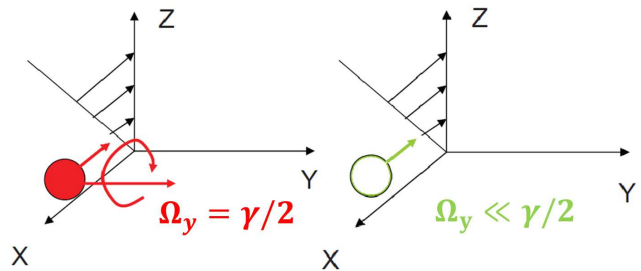}
\caption{The left sphere exhibits the behaviour of a passive particle in a shear flow. The particles roll around an axis through its centre with the half shear rate. The right sphere is an example of the behaviour of active particles that maintain a constant orientation in the shear flow.}
\label{fig:5}
\end{figure}

%It has been proposed that gravity should have a significant effect on the effective viscosity of a suspension of CR that are bottom-heavy\cite{Pedley.1992,Ishikawa.2007}.
%The centre of mass of the CR is not located at the geometrical centre, and this buoyancy was supposed to result in a persistent orientation in the gravitational field.

The difference between a passive particle that rotates with $\Omega_y=\dot\gamma/2$ and an active particle that maintains its orientation is shown in fig. \ref{fig:5}. It is obvious that the persistence of orientation leads to an additional shear stress. Previous studies \cite{Pedley.1992,Ishikawa.2007} theoretically showed that if the cells are sensitive to gravity (named bottom-heavy squirmers  due to the off-centred centre of mass), the orientation of the vorticity to the gravity field would influence the rheological measurements. If the gravitational field is perpendicular to the vorticity, the effective viscosity is increased; however, if the gravity and vorticity fields are parallel, the effective viscosity should not be affected. This is not in agreement with our observations since the strain of cells used in our work is not sensitive to gravity. Complementary experiments have shown that the random walk of the strain of CR used in this study is indeed not determined by gravity. The velocity distribution of the microswimmers is similar in the flow and gradient directions, as measured by particle tracking microscopy. Therefore, our results cannot be explained with this model. These data confirm that the effect of motility on the rheology of microswimmer suspensions is not a simple consequence of the orientation of the gravity field.

The breast-stroke-like propulsion of CR leads to the development of several models to describe the behaviour of such microswimmers, especially in shear flow.

 \revision{Since CR have a slight nonspherical shape, it is first instructive
to evaluate the contribution of ellipticity of active swimmers to the intrinsic
viscosity $\alpha=2.5+\delta$. The analytical correction $\delta$
is due to the presence of the nonspherical swimmers ($\delta<0$ for pushers, and
$\delta>0$ for pullers); it has been derived analytically by Haines
et al. \cite{Haines.2009} for a shear flow and is such that:
\begin{equation}
\label{eq:6}
\delta=\frac{27vbDN(5\lambda^{2}+10\lambda+2)}{10a^{2}(36D^{2}+\dot{\gamma}^{2})(1+\lambda)^{4}}\epsilon+\mathcal{O}(\epsilon^{2})
\end{equation}
where $v\approx50\mu m s^{-1}$ is the velocity of the CR, $a\approx7\,\mu m$
and $b\approx7.7\,\mu m$ the ellipsoidal lengths, yielding $\epsilon=b/a-1\approx0.1$.
$(1+\lambda)b$ is the typical distance from the body centre where the volume forces
are applied and we choose $\lambda\approx0.75$.
The force that is exerted by the flagella is $f=6\pi\eta bNv$, $N$ being a scalar function which rescales the drag coefficient and
following ref. \cite{Haines.2009} we use $N=0.5$. The shear rate is $\dot{\gamma}=5s^{-1}$.
Finally the rotational diffusion for algae has been estimated by Pedley
and Kessler \cite{Pedley.1992} $D\approx0.067\, s^{-1}$
(for comparison the thermal rotational diffusion coefficient for a
$10\mu m$ diameter sphere is $0.0013\, s^{-1}$). With these parameters, we find
$\delta\approx0.0037$ which is much below our experimental value
$\delta\approx2.0$. We can thus conclude that the
geometry cannot explain the observed changes in effective viscosity.}

 Another approach is the force dipole model, which proposes that the force vector, F, points along the direction of movement and that each flagellum exerts a force \textendash{}F/2. If a fixed CR is introduced into shear flow, a tumbling movement will be observed; however, a living CR is known to withstand shear flow below $10 Hz$ \cite{Rafai.2010} and to align its force vector in a constant angle to the shear stress. In a previous study \cite{Drescher.2010}, the flow field of a CR was measured directly in a fluid at rest and compared to the computed flow field of a three- or two-Stokeslet model. These results indicate that the simple puller-type description for CR is only valid at distances larger than seven times the diameter of the CR. In addition, shear flow might affect the flow field around a swimmer; thus, it is also important to study the microscopic behaviour under shear.

\section{Conclusion}

We have studied the macroscopic accessible effective viscosity of an active suspension composed of living and fixed \textit{Chlamydomonas reinhardtii}. It has been shown that for both cone-plate and Taylor-Couette geometries, the effective viscosity of suspensions of CR increases identically as a function of the volume fraction. The dependency of the viscosity on the concentration can be described by the Krieger-Dougherty law for both geometries with identical intrinsic viscosities. Microscopic measurements of the distribution of swimming directions confirmed these findings. These results led to the conclusion that the pure geometrical aspects and the inner structure of CR {[}its centre of mass eccentricity (fig. \ref{fig:5}){]} i.e., gravitational effects, are not sufficient to describe this viscosity increase from a macroscopic point of view. The change in viscosity may or may not be explained by preferential orientation of swimmers in shear flow. But it is clear that if there is a preferential orientation in our case it is not due to bottom-heaviness nor due to elongation. We think that the motility generated by the breast strokes of the flagella have to be taken into account to develop a realistic model for the description of CR in shear flow; this model is currently being developed by our teams.

\acknowledgments
This work was funded by the Graduate School 1276 (DFG). SR and PP
are funded by ANR MICMACSWIM.

\bibliographystyle{eplbib}

\end{document}